\documentclass[prl,aps,twocolumn,showpacs,letterpaper,floatfix,superscriptaddress]{revtex4-1}
\usepackage{graphicx} 
\usepackage{hyperref}
\usepackage[percent]{overpic}
\usepackage{rotating}
\usepackage{color}
\begin{document}

\title{Zigzag Phase Transition in Quantum Wires}
\author{Abhijit C. Mehta}
\affiliation{Department of Physics, Duke University, Box 90305, Durham, North Carolina, 27708-0305, USA}
\author{C. J. Umrigar}
\affiliation{Laboratory of Atomic and Solid State Physics, Cornell University, Ithaca, New York, 14853,USA}
\author{Julia S. Meyer}
\affiliation{SPSMS, UMR-E 9001 CEA/UJF-Grenoble 1, INAC, Grenoble, F-38054, France}
\author{Harold U. Baranger}
\affiliation{Department of Physics, Duke University, Box 90305, Durham, North Carolina, 27708-0305, USA}
\date{May 14, 2013} 

\begin{abstract}
We study the quantum phase transition of interacting electrons in quantum wires
from a one-dimensional (1D) linear configuration to a quasi-1D zigzag
arrangement using quantum Monte Carlo methods. As the density increases from its
lowest values, first, the electrons form a linear Wigner crystal; then, the
symmetry about the axis of the wire is broken as the electrons order in a
quasi-1D zigzag phase; and, finally, the electrons form a disordered liquid-like
phase. 
We show that the linear to zigzag phase transition is not destroyed by the strong
quantum fluctuations present in narrow wires;
it has characteristics which are qualitatively
different from the classical transition. 
\end{abstract}

\maketitle{}

Interacting one-dimensional (1D) systems have been a fruitful field of
study in both condensed matter and atomic physics~\cite{giamarchi_2003_book,imambekov_2012_rmp}. 
Experiments on 
semiconductor 
quantum wires and carbon nanotubes, for instance,
have yielded a rich set of data on the 1D electron gas over the past two decades
\cite{deshpande_2010_nature}. Linear ion traps, on the other hand, provide 
new systems
for studying fundamental 1D physics
\cite{bloch_2005_natphys} as well as potential platforms for 
quantum computing~\cite{home_2009_science} and quantum simulation~\cite{blatt_2012_natphys}.  
These experimental systems are not truly 1D, of course, 
and the transition from 1D to higher-dimensional behavior is of both practical
and theoretical interest.  Here we study the first stage in such a transition:
the 
change 
from a 1D linear arrangement of particles to a quasi-1D
zigzag configuration, and then to a liquid state at higher densities.

\begin{figure}[t]
\includegraphics[width=\columnwidth]{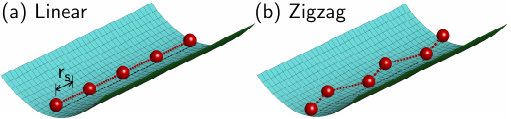}
\vspace*{-0.15in}
\caption{
(a) Electrons (red spheres) confined by a harmonic potential 
form a linear Wigner crystal. 
(b) As electron density increases, symmetry about the wire axis breaks 
and the electrons form a zigzag structure.}
\label{fig:wiremockup}
\end{figure}

At low densities, electrons confined to 1D by a transverse 
harmonic potential form a linear Wigner crystal 
~\cite{schulz_1993_prl,giamarchi_2003_book,shulenburger_2008_prb, [{For a review, see }] meyer_2009_jphys},
as illustrated in Fig.\,\ref{fig:wiremockup}(a). As the electron density is
increased (or the harmonic confinement relaxed), the Coulomb repulsion between
particles becomes comparable to the confining potential. The linear crystal 
buckles at a critical value of the electron density
~\cite{chaplik_1980_jetp,schiffer_1993_prl,piacente_2004_prb,piacente_2010_prb,galvanmoya_2011_prb}, 
breaking the symmetry about the longitudinal axis and forming a zigzag structure,
as depicted in Fig.\,\ref{fig:wiremockup}(b). This system has been studied theoretically
in both the weakly and strongly interacting limits
~\cite{meyer_2007_prl, meyer_2009_jphys, sitte_2009_prl, meng_2011_prb}: the
zigzag transition is predicted to be an Ising-type quantum phase transition in
the strongly interacting limit~\cite{meyer_2007_prl}, 
whereas at weak coupling, the critical exponents are non-universal
~\cite{meyer_2007_prl,meng_2011_prb}.  
Furthermore, the linear and zigzag
phases are expected to have only one gapless excitation mode, which corresponds
to longitudinal sliding of the crystal. At higher densities, the integrity of
the zigzag structure is destroyed, and a second mode becomes gapless.  Quasi-1D lattice 
structures were noted in numerical calculations on wires with weak confinement~\cite{welander_2010_prb}. Experimentally, evidence for a coupled two-row 
structure has been observed in the conductance of quantum wires fabricated in
GaAs/AlGaAs heterostructures~\cite{hew_2009_prl, smith_2009_prb}.

Atomic systems provide an alternative to electrons for studying
zigzag physics. Ions in linear traps, for instance, have been observed to undergo a zigzag
transition~\cite{birkl_1992_nature,enzer_2000_prl}. Though these experiments
are understood purely classically~\cite{schiffer_1993_prl},
recent theoretical work suggests that the quantum zigzag transition may be
experimentally accessible in ion trap systems
~\cite{shimshoni_2011_prl,gong_2010_prl,delcampo_2010_prl}. Dilute dipolar gases 
also undergo a related transition, studied in recent theoretical and numerical work
~\cite{astrakharchik_2008_pra,ruhman_2012_prb}.

To connect to experiments in electronic systems, the behavior at intermediate
interaction strength is key.  
We observe the zigzag transition at intermediate interaction strength, 
characterizing it through both the pair density and a correlation function of the zigzag order. 
By studying the long-range zigzag correlations, we demonstrate that
the quantum phase transition occurs at parameters relevant to quantum wire
experiments and is qualitatively different from the classical transition.

The quantum wire consists of $N$ spinless electrons with Coulomb interactions
confined to the circumference of a ring of radius $R$ by a harmonic potential
with frequency $\omega$, as described by the two-dimensional Hamiltonian  
\begin{equation}
\label{eq:ringhamiltonian}
H = -\frac{1}{2} \sum_{i=1}^N \nabla_i^2 
+ \frac{1}{2}\sum_{i=1}^N \omega^2 (r_i -R)^2 
+ \sum_{i < j \leq N} \frac{1}{|\mathbf{r}_i - \mathbf{r}_j|}\mathrm{.}
\end{equation}
We use effective atomic units: the effective mass $m^*$, the electric
charge $e$, the dielectric constant $\epsilon$, and $\hbar$ are all set to 1. In
GaAs, the effective Bohr radius $a_0^* = \hbar^2 \epsilon / m^* e^2$ is 9.8 nm,
and the energy scale---the effective Hartree $H^* = e^2/\epsilon a_0^*$---is
11.9 meV. Using a ring is a convenient way to impose periodic boundary
conditions on the quantum wire.
The middle of the (effective) wire is then defined by the average radial coordinate, $\bar{r} \equiv \int \!dr\, r\, n(r)$, and the longitudinal coordinate along the wire can be taken to be the 2D angular coordinate $\theta$. 

Two length scales in our system are of particular interest: 
the Wigner-Seitz radius $r_s \equiv 1/2n_{\rm 1D}$ 
(where $n_{\rm 1D}$ is the linear density),
and the length scale $r_0$ at which the Coulomb interaction between 
neighboring electrons becomes comparable to the harmonic confinement,
\begin{equation}
\label{eq:romega}
r_0 \equiv \sqrt[3]{\frac{2 e^2}{\epsilon m^* \omega^2}} \mathrm{.}
\end{equation}
The zigzag transition occurs when the length scales $r_s$ and $r_0$ 
become comparable~\cite{meyer_2009_jphys}. 
The classical transition has been studied for electrons in liquid helium
and for ion trap systems
~\cite{chaplik_1980_jetp,schiffer_1993_prl,piacente_2004_prb,piacente_2010_prb,galvanmoya_2011_prb}.  
For our system, calculation of the classical 
critical point is straightforward,
and the width between rows 
scales as $\sqrt{n-n_{\rm critical}}$~\cite{meyer_2009_jphys}.
The classical description is valid when the effective Bohr radius is much smaller
than the interparticle separation at the zigzag transition; i.e., when $r_0 \gg 1$.
Note that for large $r_0$ (small $\omega$, thus wider wires)
the electrons are effectively more strongly interacting. 
For smaller values of $r_0$ (narrower wires), quantum fluctuations 
play an important role. 
We focus on two values of the confinement, $\omega=0.1$ and $0.6$, which 
correspond to $r_0 = 5.9$ and $1.8$, respectively, because they correspond 
to experimentally measured parameters in GaAs/AlGaAs quantum wires
~\cite{rossler_2011_newjphys}.
At $\omega = 0.6$ in particular, quantum effects play an important role 
in the transition since $r_0$ is close to $1$.

We calculate the ground state properties of our system using 
Quantum Monte Carlo (QMC) techniques~\cite{foulkes_2001_rmp,umrigar_1993_jchemphys}. 
In the first step of our QMC calculation, Variational Monte Carlo (VMC), 
we minimize the variational energy of a Slater-Jastrow type trial wave function
$\Psi_T(\mathbf{R}) = J(\mathbf{R})D(\mathbf{R})$~\cite{guclu_2005_prb}, 
using methods described in~\cite{umrigar_2005_prl} and~\cite{umrigar_2007_prl}. 
We consider three qualitatively different types of single-particle orbitals
to build the Slater determinant $D(\mathbf{R})$---
localized floating gaussians~\cite{guclu_2008_prb}, 
planewaves, and orbitals from density functional calculations---and 
use the type yielding the lowest variational energy at a given density.  
After optimizing the variational parameters, we use 
Diffusion Monte Carlo (DMC) to project the trial wavefunction onto the
fixed-node approximation of the ground state~\cite{umrigar_1993_jchemphys,foulkes_2001_rmp}.
The fixed-node DMC wavefunction is the lowest-energy state 
with the nodes of the trial wavefunction $\Psi_T$,
and its energy is an upper bound on the true ground state energy.  
We use an extrapolated estimator $\langle\hat O \rangle_{\rm QMC} =
2\langle\hat O \rangle_{\rm DMC} - \langle\hat O \rangle_{\rm VMC}$
to calculate observables $\hat O$ that do not commute with
the Hamiltonian~\cite{foulkes_2001_rmp}.

\begin{figure}[thb]
\vspace*{-0.15in}
\includegraphics[width=\columnwidth]{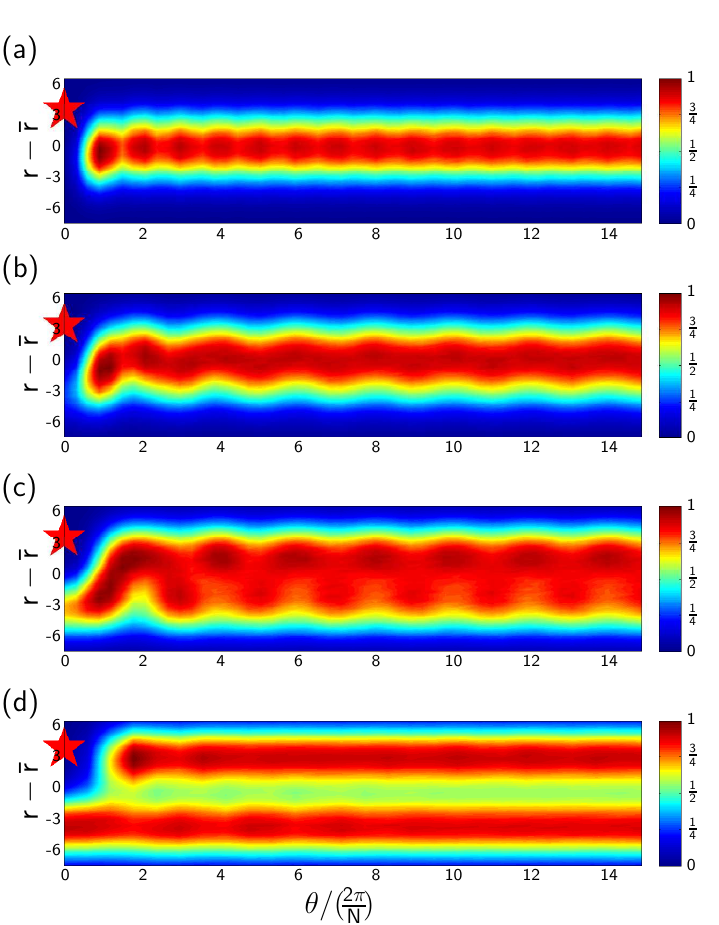}
\vspace*{-0.15in}
\caption{Pair density, 
$\langle \hat\rho(r,\theta)\hat\rho(r_\star,0) \rangle 
/\langle \hat\rho(r_\star,0)\rangle$, 
for electrons at progressively higher densities in
a wire with $\omega=0.1$. The red stars mark $r_\star$. 
($r$ is plotted relative to $\bar{r}$ in units 
of the effective Bohr radius $a_0^*$,
and $\theta$ is plotted in units of the interparticle spacing $2\pi/N$.)
(a)~$r_s = 4.0$, $N = 30$. At low densities, the electrons form a linear Wigner crystal. 
(b)~$r_s = 3.6$, $N = 30$. As the density increases, a zigzag structure forms. 
(c) $r_s = 3.0$, $N = 30$. The amplitude of the zigzag increases.
(d) $r_s = 2.0$, $N = 60$. At higher densities, the zigzag structure 
is destroyed.
The color scale shows the pair density relative to the maximum value 
for each plot.
Only a portion of the full periodic system is shown.
}
\label{fig:pairden}
\end{figure}

The pair density, defined as
$\left\langle \hat\rho(r,\theta) \hat\rho(r_\star,0)\right\rangle/
\langle \hat\rho(r_\star,0)\rangle$ where $\hat\rho(r,\theta)$ is the local density operator and $(r_\star,0)$ is the location of a fixed electron,   
is a key microscopic quantity that allows direct visualization
of the system's quasi-1D nature.  Fig.\,\ref{fig:pairden} 
shows the pair densities in different phases 
of a quantum wire with $\omega \!=\! 0.1$.
At $r_s = 4.0$---the ``linear phase"---modulations in the pair density indicate 
that the electrons are quasi-localized in a linear arrangement, as  
observed in previous QMC calculations~\cite{casula_2006_prb,shulenburger_2008_prb}.
Fig.\,\ref{fig:pairden}(a) shows that this linear ordering 
persists even when $r_\star$ deviates significantly 
from the center of the wire: there is, of course, some short-range zigzag correlation 
in response to the off-axis electron, but at larger distances the arrangement is linear.  
We observe this linear phase until the
density increases past $r_s = 3.79$, where the zigzag transition occurs; 
this value is quite close to that for the classical transition, 
$r_s^{\rm class.} = 3.75$ for $\omega=0.1$.
Fig.\,\ref{fig:pairden}(b) shows the system in the ``zigzag phase" at 
$r_s = 3.6$:  
the electrons are arranged in a long-range zigzag pattern.
Beyond the transition, the amplitude of the zigzag structure continues to
increase, reaching a maximum value near $r_s = 3.0$, shown in
Fig.\,\ref{fig:pairden}(c).

\begin{figure}[tb]
\includegraphics[width=\columnwidth]{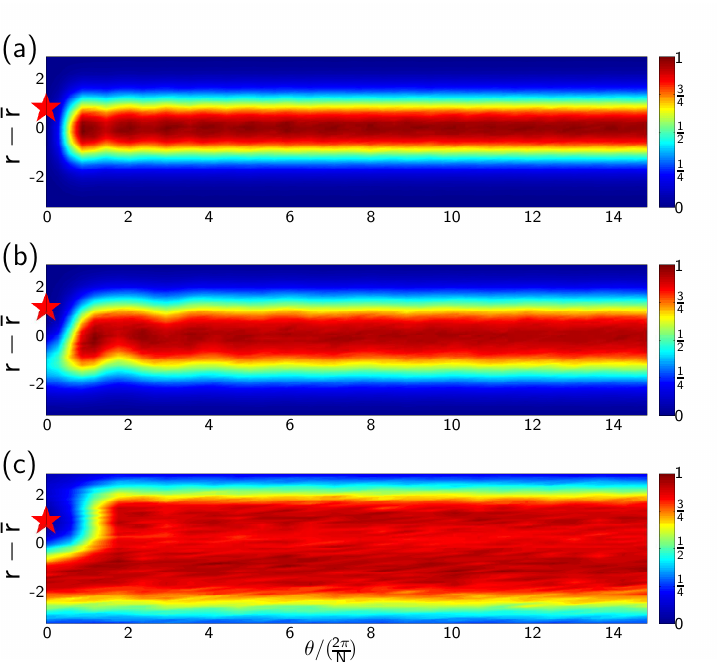}
\vspace*{-0.15in}
\caption{Pair densities 
for electrons in 
a wire with $\omega=0.6$. The red stars mark $r_\star$. 
($r$ is plotted relative to $\bar{r}$,
and $\theta$ is plotted in units of the interparticle spacing $2\pi/N$.)
(a) $r_s = 1.5, N = 30$. At low densities, electrons in a quantum wire form a linear Wigner crystal. 
(b) $r_s = 1.3, N = 30$. As the density increases, a zigzag structure forms, though quantum
fluctuations smear our correlations in the pair density. 
(c) $r_s = 0.5, N = 60$. At higher densities, the zigzag structure 
vanishes.
The color scale shows the pair density relative to the maximum value 
for each plot.
Only a portion of the full periodic system is shown.
}
\label{fig:pairdenw06}
\end{figure}

We now turn to the case with stronger quantum fluctuations, $\omega = 0.6$, 
where $r_0$ is on the same scale as the effective Bohr radius ($r_0 = 1.8$). 
At $r_s = 1.5$, Fig.\,\ref{fig:pairdenw06}(a) shows the system in
the linear phase, but modulations in the pair density are much smaller than at
$\omega=0.1$.  The zigzag transition occurs between $r_s = 1.4$ and $r_s = 1.45$; this 
deviates significantly from the classical transition point, 
$r_s^{\rm class} = 1.19$ for $\omega = 0.6$.
Fig.\,\ref{fig:pairdenw06}(b) shows the system in the zigzag phase at $r_s = 1.3$;
quantum fluctuations have smeared out correlations in the pair density.
We shall now show, however, that there are strong zigzag correlations present 
despite the rather weak features in the pair density.

\begin{figure}[tb]
\includegraphics[width=\columnwidth]{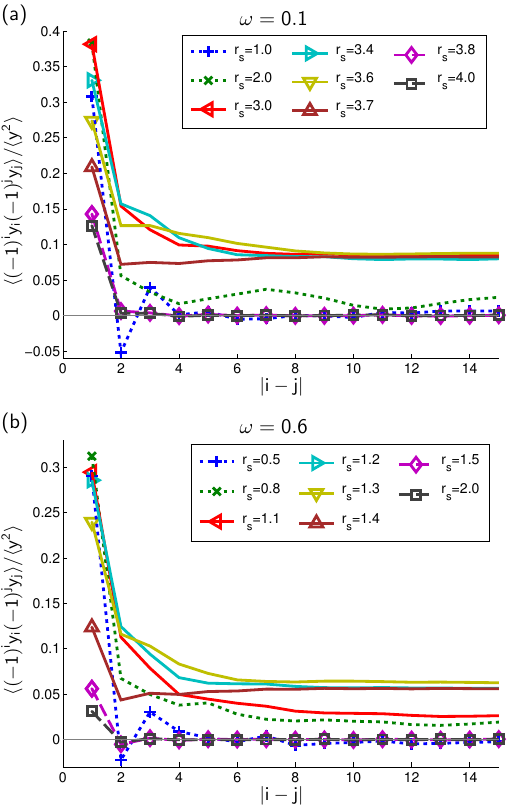}
\vspace*{-0.15in}
\caption{The zigzag correlation function, normalized by the average
wire width;
$C_{zz}(|i-j|)/\langle y^2 \rangle = \langle (-1^i) y_i (-1^j) y_j\rangle / \langle y^2 \rangle$, 
plotted for various values of $r_s$ at (a) $\omega = 0.1$ 
and (b) $\omega = 0.6$ for $N=30$ electrons. ($y \equiv r - \bar{r}$ is
measured in units of the effective Bohr radius $a_0^*$.)
In the linear phase (dashed lines), $C_{zz}$ decays to $0$; in the
zigzag phase (solid lines), $C_{zz}$ saturates to a finite value, 
indicating long-range zigzag order.  In the liquid phase (dotted lines)
$C_{zz}$ again decays.
[Note that since our system is periodic, $C_{zz}(k) = C_{zz}(N-k)$, thus
$C_{zz}(k)$ must be flat at $k=N/2$.]
}
\label{fig:zzcorrij}
\end{figure}

To characterize the long-range zigzag order more quantitatively, we first number the
electrons along the wire axis (i.e., by increasing $\theta$).  We can then define a
zigzag correlation function 
\begin{equation}
\label{eq:zzcorr}
C_{zz}(|i-j|) \equiv \left\langle (-1)^i y_i (-1)^j y_j\right\rangle
\mathrm{,}
\end{equation}
where $y \equiv r - \bar{r}$ denotes the transverse coordinate 
(for our ring geometry, the radial position $r$ relative to the mean $\bar{r}$). 
This correlation function corresponds to the field $(-1)^i y_i$ that orders in the zigzag
state~\cite{meyer_2007_prl};
$C_{zz}(|i-j|)$ indicates how strongly the $i^{\rm th}$ and $j^{\rm th}$ electrons are locked
in a zigzag pattern as a function of the number of intervening electrons.
Note that $C_{zz}$ is similar to a staggered spin correlation function 
for an antiferromagnetic system.

Fig.\,\ref{fig:zzcorrij} shows the zigzag correlation function, normalized by
the mean squared wire width $\langle y^2 \rangle$, plotted
at several values of $r_s$ for $\omega=0.1$ [Fig.\,\ref{fig:zzcorrij}(a)]
and $\omega = 0.6$ [Fig.\,\ref{fig:zzcorrij}(b)]. 
In the linear phase ($r_s = 3.8, 4.0$ for $\omega = 0.1$; $r_s = 1.5,2.0$ for $\omega = 0.6$), 
there is no long-range zigzag order,
and $C_{zz}$ decays to $0$ within a few inter-particle spacings.  
In the zigzag phase ($r_s = 3.0,\ldots,3.7$ for $\omega=0.1$ and 
$r_s = 1.1,\ldots,1.4$ for $\omega=0.6$), $C_{zz}$ saturates 
to a finite value. 
At $\omega = 0.6$, long-range zigzag order is present even 
in the absence of strong long-range positional order along the axis of the wire.  
This is possible because the zigzag order is not local (tied to the coordinate
along the wire axis), but rather depends non-locally on the numbering of the electrons
along the wire.
This demonstrates that the zigzag transition occurs in the quantum regime.

When we increase the density further, the zigzag correlation function
again decays, indicating that the zigzag structure disappears. 
This is visible in Fig.\,\ref{fig:zzcorrij} at higher densities 
($r_s \!=\! 1.0,2.0$ for $\omega \!=\!0.1$, and $r_s \!=\! 0.5,0.8$ for $\omega \!=\! 0.6$). 
At these higher densities, the pair density plots show little structure,
indicating that the positional order has been lost,
as seen in 
Fig.\,\ref{fig:pairden}(d) at $\omega\!=\!0.1$, $r_s\!=\!2.0$, and in
Fig.\,\ref{fig:pairdenw06}(c) at $\omega\!=\!0.6$, $r_s \!=\! 0.5$.
We plot results from a larger system size in these cases, $N\!=\!60$,
to lessen the effects of our ring geometry; on the right hand side of the strips in Figs.\,\ref{fig:pairden} and \ref{fig:pairdenw06}, the small difference between the inner and outer edge ($r - \bar{r} \!<\! 0$ or $>\!0$, respectively) demonstrates the small effect of annularity. 
Two rows are visible in the pair density, but there are no strong
zigzag correlations. We identify this liquid-like phase with the two-gapless-mode phase 
described in~\cite{meyer_2007_prl}.   

There are a number of unusual features in $C_{zz}$. 
First, for values of $r_s$ close to the zigzag transition 
(e.g., $r_s \!= 3.7$ at $\omega \!= 0.1$, and $r_s \!= 1.4$ at $\omega \!= 0.6$), 
$C_{zz}$ decreases sharply before saturating, while the decay is more gradual for
smaller $r_s$.  
Also, at the highest densities, $C_{zz}$ shows anti-zigzag ordering 
at $|i\!-\!j|\!=\!2$ \footnote{  
This is due to the decoupling of the two rows;
at high densities, there are likely to be configurations where two ``adjacent''
(ordered by their $\theta$ coordinate) electrons are in the same row, and the next
electron is in the opposite row.  (I.e., the $|i-j|=0$ and $|i-j|=1$ electrons
are in the same row, and the $|i-j|=2$ electron is in the opposite row, or
both the $|i-j|=1$ and $|i-j|=2$ electrons are in the row opposite the $|i-j|=0$
electron.)  Such configurations will give a negative contribution to $C_{zz}(|i-j|=2)$.
}.
Finally, we see oscillations in $C_{zz}$ in the phase where the zigzag structure is destroyed 
($r_s \!=\! 1.0,2.0$ at $\omega \!=\! 0.1$, and $r_s \!=\! 0.5,0.8$ at $\omega \!=\! 0.6$). Since we observe longer wavelengths at larger $N$ ($N\!=\!60$), these oscillations appear to be caused by finite-size effects; the main features of $C_{zz}$ discussed above, however, are not changed in the larger system. 

\begin{figure}[t]
\includegraphics[width=\columnwidth]{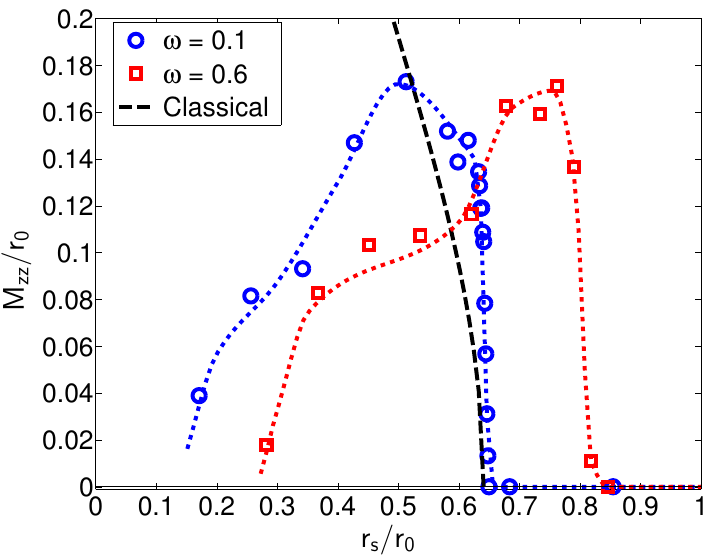}
\vspace*{-0.15in}
\caption{The ``Zigzag Amplitude" order parameter $M_{zz}$, 
as a function of $r_s$ at $\omega = 0.1$ (blue circles) and $\omega = 0.6$ (red squares).
As $r_s$ is decreased beyond the critical value and the system enters the zigzag regime, 
$M_{zz}$ increases sharply; at lower $r_s$, $M_{zz}$ decreases gradually as the
system enters the liquid regime.
The behavior deviates significantly from the $\sqrt{n-n_{\rm critical}}$ behavior for classical electrons
(black dashed line). 
Dotted lines are a guide to the eye.
(Lengths are scaled by $r_0$. We note that there are points for
both values of $\omega$ where $M_{zz}$ seems to be lower than one would
expect by drawing a curve through the other points; we believe that this
is a systematic error from the VMC optimization step of our QMC calculation.) 
}
\label{fig:zza_corrfun_scaled}
\end{figure}

The long-range value of the zigzag correlation yields 
the order parameter of the phase transition, $M_{zz}$. We estimate 
$M^2_{zz}$ by averaging over the long-range part of the correlation,
\begin{equation}
\label{eq:mzz}
M^2_{zz} \approx \langle C_{zz}(|i-j|)\rangle_{|i-j|>N/4} \mathrm{.}
\end{equation}
$M_{zz}$ is related to the amplitude of the zigzag structure---it
is
half the width between the two zigzag rows.  
Fig.\,\ref{fig:zza_corrfun_scaled} shows 
the order parameter at $\omega = 0.1$ and 
$0.6$ as a function of $r_s$ (scaled in units of $r_0$). 
The fine scale non-monotonic behavior in both data sets is an indication of the error coming from the VMC optimization step. 

$M_{zz}$ increases sharply at the transition from the linear 
phase to the zigzag phase.
As with
the pair densities and correlation functions, $M_{zz}$ shows the same generic
behavior at $\omega = 0.6$ as at $0.1$---behavior consistent with that of a continuous phase transition.
The transition occurs at a scaled density close to the classical
value for $\omega = 0.1$ but at a considerably lower scaled density for $\omega = 0.6$.  In both cases, the behavior near the transition differs qualitatively 
from the classical case~\cite{meyer_2009_jphys}, highlighting its quantum nature.
As the system evolves at higher density from the zigzag to the liquid-like phase, 
$M_{zz}$ decreases gradually.

In conclusion, we have demonstrated that the linear to zigzag transition occurs 
at parameters relevant to experiments in quantum wires.
A transition to a phase with long-range zigzag correlations
occurs even in narrow wires where large quantum fluctuations smear 
out density correlations;
the quantum phase transition in these wires differs substantially from the classical case.
Future work will examine other signatures of this transition;  
in addition, the role of spin and finite temperature at intermediate 
interaction strength remains unexplored.

We thank J. Anders, A. D. G\"{u}\c{c}l\"{u}, K. A. Matveev, J. Shumway, and M. Pepper 
for helpful discussions, and H. Jiang for providing the density functional code. 
This work was supported in part by the U.S. DOE, Materials Sciences and Engineering, 
DE-SC0005237 (A.C.M. and H.U.B), the U.S. NSF, DMR-0908653 (C.J.U.), and
EU-FP7 Marie Curie IRG (J.S.M.).
A.C.M. was partially supported by an E. Bayard Halsted Fellowship from Duke University.
A portion of the calculations were done on the Open Science Grid,
which is supported by the NSF and DOE Office of Science.
J.S.M. and H.U.B. thank the
Fondation Nanosiences de Grenoble for facilitating the exchange between Grenoble and Duke.

\bibliography{zigzag_short_paper}

\begin{thebibliography}{36}%
\makeatletter
\providecommand \@ifxundefined [1]{%
 \@ifx{#1\undefined}
}%
\providecommand \@ifnum [1]{%
 \ifnum #1\expandafter \@firstoftwo
 \else \expandafter \@secondoftwo
 \fi
}%
\providecommand \@ifx [1]{%
 \ifx #1\expandafter \@firstoftwo
 \else \expandafter \@secondoftwo
 \fi
}%
\providecommand \natexlab [1]{#1}%
\providecommand \enquote  [1]{``#1''}%
\providecommand \bibnamefont  [1]{#1}%
\providecommand \bibfnamefont [1]{#1}%
\providecommand \citenamefont [1]{#1}%
\providecommand \href@noop [0]{\@secondoftwo}%
\providecommand \href [0]{\begingroup \@sanitize@url \@href}%
\providecommand \@href[1]{\@@startlink{#1}\@@href}%
\providecommand \@@href[1]{\endgroup#1\@@endlink}%
\providecommand \@sanitize@url [0]{\catcode `\\12\catcode `\$12\catcode
  `\&12\catcode `\#12\catcode `\^12\catcode `\_12\catcode `\%12\relax}%
\providecommand \@@startlink[1]{}%
\providecommand \@@endlink[0]{}%
\providecommand \url  [0]{\begingroup\@sanitize@url \@url }%
\providecommand \@url [1]{\endgroup\@href {#1}{\urlprefix }}%
\providecommand \urlprefix  [0]{URL }%
\providecommand \Eprint [0]{\href }%
\providecommand \doibase [0]{http://dx.doi.org/}%
\providecommand \selectlanguage [0]{\@gobble}%
\providecommand \bibinfo  [0]{\@secondoftwo}%
\providecommand \bibfield  [0]{\@secondoftwo}%
\providecommand \translation [1]{[#1]}%
\providecommand \BibitemOpen [0]{}%
\providecommand \bibitemStop [0]{}%
\providecommand \bibitemNoStop [0]{.\EOS\space}%
\providecommand \EOS [0]{\spacefactor3000\relax}%
\providecommand \BibitemShut  [1]{\csname bibitem#1\endcsname}%
\let\auto@bib@innerbib\@empty
\bibitem [{\citenamefont {Giamarchi}(2003)}]{giamarchi_2003_book}%
  \BibitemOpen
  \bibfield  {author} {\bibinfo {author} {\bibfnamefont {T.}~\bibnamefont
  {Giamarchi}},\ }\href@noop {} {\emph {\bibinfo {title} {Quantum Physics in
  One Dimension}}}\ (\bibinfo  {publisher} {Clarendon Press},\ \bibinfo {year}
  {2003})\BibitemShut {NoStop}%
\bibitem [{\citenamefont {Imambekov}\ \emph {et~al.}(2012)\citenamefont
  {Imambekov}, \citenamefont {Schmidt},\ and\ \citenamefont
  {Glazman}}]{imambekov_2012_rmp}%
  \BibitemOpen
  \bibfield  {author} {\bibinfo {author} {\bibfnamefont {A.}~\bibnamefont
  {Imambekov}}, \bibinfo {author} {\bibfnamefont {T.~L.}\ \bibnamefont
  {Schmidt}}, \ and\ \bibinfo {author} {\bibfnamefont {L.~I.}\ \bibnamefont
  {Glazman}},\ }\href {\doibase 10.1103/RevModPhys.84.1253} {\bibfield
  {journal} {\bibinfo  {journal} {Rev. Mod. Phys.}\ }\textbf {\bibinfo {volume}
  {84}},\ \bibinfo {pages} {1253} (\bibinfo {year} {2012})}\BibitemShut
  {NoStop}%
\bibitem [{\citenamefont {Deshpande}\ \emph {et~al.}(2010)\citenamefont
  {Deshpande}, \citenamefont {Bockrath}, \citenamefont {Glazman},\ and\
  \citenamefont {Yacoby}}]{deshpande_2010_nature}%
  \BibitemOpen
  \bibfield  {author} {\bibinfo {author} {\bibfnamefont {V.~V.}\ \bibnamefont
  {Deshpande}}, \bibinfo {author} {\bibfnamefont {M.}~\bibnamefont {Bockrath}},
  \bibinfo {author} {\bibfnamefont {L.~I.}\ \bibnamefont {Glazman}}, \ and\
  \bibinfo {author} {\bibfnamefont {A.}~\bibnamefont {Yacoby}},\ }\href
  {\doibase 10.1038/nature08918} {\bibfield  {journal} {\bibinfo  {journal}
  {Nature}\ }\textbf {\bibinfo {volume} {464}},\ \bibinfo {pages} {209}
  (\bibinfo {year} {2010})}\BibitemShut {NoStop}%
\bibitem [{\citenamefont {Bloch}(2005)}]{bloch_2005_natphys}%
  \BibitemOpen
  \bibfield  {author} {\bibinfo {author} {\bibfnamefont {I.}~\bibnamefont
  {Bloch}},\ }\href {\doibase 10.1038/nphys138} {\bibfield  {journal} {\bibinfo
   {journal} {Nature Physics}\ }\textbf {\bibinfo {volume} {1}},\ \bibinfo
  {pages} {23} (\bibinfo {year} {2005})}\BibitemShut {NoStop}%
\bibitem [{\citenamefont {Home}\ \emph {et~al.}(2009)\citenamefont {Home},
  \citenamefont {Hanneke}, \citenamefont {Jost}, \citenamefont {Amini},
  \citenamefont {Leibfried},\ and\ \citenamefont
  {Wineland}}]{home_2009_science}%
  \BibitemOpen
  \bibfield  {author} {\bibinfo {author} {\bibfnamefont {J.~P.}\ \bibnamefont
  {Home}}, \bibinfo {author} {\bibfnamefont {D.}~\bibnamefont {Hanneke}},
  \bibinfo {author} {\bibfnamefont {J.~D.}\ \bibnamefont {Jost}}, \bibinfo
  {author} {\bibfnamefont {J.~M.}\ \bibnamefont {Amini}}, \bibinfo {author}
  {\bibfnamefont {D.}~\bibnamefont {Leibfried}}, \ and\ \bibinfo {author}
  {\bibfnamefont {D.~J.}\ \bibnamefont {Wineland}},\ }\href {\doibase
  10.1126/science.1177077} {\bibfield  {journal} {\bibinfo  {journal}
  {Science}\ }\textbf {\bibinfo {volume} {325}},\ \bibinfo {pages} {1227}
  (\bibinfo {year} {2009})}\BibitemShut {NoStop}%
\bibitem [{\citenamefont {Blatt}\ and\ \citenamefont
  {Roos}(2012)}]{blatt_2012_natphys}%
  \BibitemOpen
  \bibfield  {author} {\bibinfo {author} {\bibfnamefont {R.}~\bibnamefont
  {Blatt}}\ and\ \bibinfo {author} {\bibfnamefont {C.~F.}\ \bibnamefont
  {Roos}},\ }\href {\doibase 10.1038/nphys2252} {\bibfield  {journal} {\bibinfo
   {journal} {Nature Physics}\ }\textbf {\bibinfo {volume} {8}},\ \bibinfo
  {pages} {277} (\bibinfo {year} {2012})}\BibitemShut {NoStop}%
\bibitem [{\citenamefont {Schulz}(1993)}]{schulz_1993_prl}%
  \BibitemOpen
  \bibfield  {author} {\bibinfo {author} {\bibfnamefont {H.~J.}\ \bibnamefont
  {Schulz}},\ }\href {\doibase 10.1103/PhysRevLett.71.1864} {\bibfield
  {journal} {\bibinfo  {journal} {Phys. Rev. Lett.}\ }\textbf {\bibinfo
  {volume} {71}},\ \bibinfo {pages} {1864} (\bibinfo {year}
  {1993})}\BibitemShut {NoStop}%
\bibitem [{\citenamefont {Shulenburger}\ \emph {et~al.}(2008)\citenamefont
  {Shulenburger}, \citenamefont {Casula}, \citenamefont {Senatore},\ and\
  \citenamefont {Martin}}]{shulenburger_2008_prb}%
  \BibitemOpen
  \bibfield  {author} {\bibinfo {author} {\bibfnamefont {L.}~\bibnamefont
  {Shulenburger}}, \bibinfo {author} {\bibfnamefont {M.}~\bibnamefont
  {Casula}}, \bibinfo {author} {\bibfnamefont {G.}~\bibnamefont {Senatore}}, \
  and\ \bibinfo {author} {\bibfnamefont {R.~M.}\ \bibnamefont {Martin}},\
  }\href {\doibase 10.1103/PhysRevB.78.165303} {\bibfield  {journal} {\bibinfo
  {journal} {Phys. Rev. B}\ }\textbf {\bibinfo {volume} {78}},\ \bibinfo
  {pages} {165303} (\bibinfo {year} {2008})}\BibitemShut {NoStop}%
\bibitem [{\citenamefont {Meyer}\ and\ \citenamefont
  {Matveev}(2009)}]{meyer_2009_jphys}%
  \BibitemOpen
  \bibfield  {author} {\bibinfo {author} {\bibfnamefont {J.~S.}\ \bibnamefont
  {Meyer}}\ and\ \bibinfo {author} {\bibfnamefont {K.~A.}\ \bibnamefont
  {Matveev}},\ }\href {\doibase 10.1088/0953-8984/21/2/023203} {\bibfield
  {journal} {\bibinfo  {journal} {J. Phys.: Condens. Matter}\ }\textbf
  {\bibinfo {volume} {21}},\ \bibinfo {pages} {023203} (\bibinfo {year}
  {2009})}\BibitemShut {NoStop}%
\bibitem [{\citenamefont {Chaplik}(1980)}]{chaplik_1980_jetp}%
  \BibitemOpen
  \bibfield  {author} {\bibinfo {author} {\bibfnamefont {A.~K.}\ \bibnamefont
  {Chaplik}},\ }\href
  {http://www.jetpletters.ac.ru/ps/1344/article_20292.shtml} {\bibfield
  {journal} {\bibinfo  {journal} {Pis'ma Zh. Eksp. Teor. Fiz.}\ }\textbf
  {\bibinfo {volume} {31}},\ \bibinfo {pages} {275} (\bibinfo {year} {1980})},\
  \bibinfo {note} {[JETP Lett. {\bf 31}, 252 (1980)]}\BibitemShut {NoStop}%
\bibitem [{\citenamefont {Schiffer}(1993)}]{schiffer_1993_prl}%
  \BibitemOpen
  \bibfield  {author} {\bibinfo {author} {\bibfnamefont {J.~P.}\ \bibnamefont
  {Schiffer}},\ }\href {\doibase 10.1103/PhysRevLett.70.818} {\bibfield
  {journal} {\bibinfo  {journal} {Phys. Rev. Lett.}\ }\textbf {\bibinfo
  {volume} {70}},\ \bibinfo {pages} {818} (\bibinfo {year} {1993})}\BibitemShut
  {NoStop}%
\bibitem [{\citenamefont {Piacente}\ \emph {et~al.}(2004)\citenamefont
  {Piacente}, \citenamefont {Schweigert}, \citenamefont {Betouras},\ and\
  \citenamefont {Peeters}}]{piacente_2004_prb}%
  \BibitemOpen
  \bibfield  {author} {\bibinfo {author} {\bibfnamefont {G.}~\bibnamefont
  {Piacente}}, \bibinfo {author} {\bibfnamefont {I.~V.}\ \bibnamefont
  {Schweigert}}, \bibinfo {author} {\bibfnamefont {J.~J.}\ \bibnamefont
  {Betouras}}, \ and\ \bibinfo {author} {\bibfnamefont {F.~M.}\ \bibnamefont
  {Peeters}},\ }\href {\doibase 10.1103/PhysRevB.69.045324} {\bibfield
  {journal} {\bibinfo  {journal} {Phys. Rev. B}\ }\textbf {\bibinfo {volume}
  {69}},\ \bibinfo {pages} {045324} (\bibinfo {year} {2004})}\BibitemShut
  {NoStop}%
\bibitem [{\citenamefont {Piacente}\ \emph {et~al.}(2010)\citenamefont
  {Piacente}, \citenamefont {Hai},\ and\ \citenamefont
  {Peeters}}]{piacente_2010_prb}%
  \BibitemOpen
  \bibfield  {author} {\bibinfo {author} {\bibfnamefont {G.}~\bibnamefont
  {Piacente}}, \bibinfo {author} {\bibfnamefont {G.~Q.}\ \bibnamefont {Hai}}, \
  and\ \bibinfo {author} {\bibfnamefont {F.~M.}\ \bibnamefont {Peeters}},\
  }\href {\doibase 10.1103/PhysRevB.81.024108} {\bibfield  {journal} {\bibinfo
  {journal} {Phys. Rev. B}\ }\textbf {\bibinfo {volume} {81}},\ \bibinfo
  {pages} {024108} (\bibinfo {year} {2010})}\BibitemShut {NoStop}%
\bibitem [{\citenamefont {Galv\'an-Moya}\ and\ \citenamefont
  {Peeters}(2011)}]{galvanmoya_2011_prb}%
  \BibitemOpen
  \bibfield  {author} {\bibinfo {author} {\bibfnamefont {J.~E.}\ \bibnamefont
  {Galv\'an-Moya}}\ and\ \bibinfo {author} {\bibfnamefont {F.~M.}\ \bibnamefont
  {Peeters}},\ }\href {\doibase 10.1103/PhysRevB.84.134106} {\bibfield
  {journal} {\bibinfo  {journal} {Phys. Rev. B}\ }\textbf {\bibinfo {volume}
  {84}},\ \bibinfo {pages} {134106} (\bibinfo {year} {2011})}\BibitemShut
  {NoStop}%
\bibitem [{\citenamefont {Meyer}\ \emph {et~al.}(2007)\citenamefont {Meyer},
  \citenamefont {Matveev},\ and\ \citenamefont {Larkin}}]{meyer_2007_prl}%
  \BibitemOpen
  \bibfield  {author} {\bibinfo {author} {\bibfnamefont {J.~S.}\ \bibnamefont
  {Meyer}}, \bibinfo {author} {\bibfnamefont {K.~A.}\ \bibnamefont {Matveev}},
  \ and\ \bibinfo {author} {\bibfnamefont {A.~I.}\ \bibnamefont {Larkin}},\
  }\href {\doibase 10.1103/PhysRevLett.98.126404} {\bibfield  {journal}
  {\bibinfo  {journal} {Phys. Rev. Lett.}\ }\textbf {\bibinfo {volume} {98}},\
  \bibinfo {pages} {126404} (\bibinfo {year} {2007})}\BibitemShut {NoStop}%
\bibitem [{\citenamefont {Sitte}\ \emph {et~al.}(2009)\citenamefont {Sitte},
  \citenamefont {Rosch}, \citenamefont {Meyer}, \citenamefont {Matveev},\ and\
  \citenamefont {Garst}}]{sitte_2009_prl}%
  \BibitemOpen
  \bibfield  {author} {\bibinfo {author} {\bibfnamefont {M.}~\bibnamefont
  {Sitte}}, \bibinfo {author} {\bibfnamefont {A.}~\bibnamefont {Rosch}},
  \bibinfo {author} {\bibfnamefont {J.~S.}\ \bibnamefont {Meyer}}, \bibinfo
  {author} {\bibfnamefont {K.~A.}\ \bibnamefont {Matveev}}, \ and\ \bibinfo
  {author} {\bibfnamefont {M.}~\bibnamefont {Garst}},\ }\href {\doibase
  10.1103/PhysRevLett.102.176404} {\bibfield  {journal} {\bibinfo  {journal}
  {Phys. Rev. Lett.}\ }\textbf {\bibinfo {volume} {102}},\ \bibinfo {pages}
  {176404} (\bibinfo {year} {2009})}\BibitemShut {NoStop}%
\bibitem [{\citenamefont {Meng}\ \emph {et~al.}(2011)\citenamefont {Meng},
  \citenamefont {Dixit}, \citenamefont {Garst},\ and\ \citenamefont
  {Meyer}}]{meng_2011_prb}%
  \BibitemOpen
  \bibfield  {author} {\bibinfo {author} {\bibfnamefont {T.}~\bibnamefont
  {Meng}}, \bibinfo {author} {\bibfnamefont {M.}~\bibnamefont {Dixit}},
  \bibinfo {author} {\bibfnamefont {M.}~\bibnamefont {Garst}}, \ and\ \bibinfo
  {author} {\bibfnamefont {J.~S.}\ \bibnamefont {Meyer}},\ }\href {\doibase
  10.1103/PhysRevB.83.125323} {\bibfield  {journal} {\bibinfo  {journal} {Phys.
  Rev. B}\ }\textbf {\bibinfo {volume} {83}},\ \bibinfo {pages} {125323}
  (\bibinfo {year} {2011})}\BibitemShut {NoStop}%
\bibitem [{\citenamefont {Welander}\ \emph {et~al.}(2010)\citenamefont
  {Welander}, \citenamefont {Yakimenko},\ and\ \citenamefont
  {Berggren}}]{welander_2010_prb}%
  \BibitemOpen
  \bibfield  {author} {\bibinfo {author} {\bibfnamefont {E.}~\bibnamefont
  {Welander}}, \bibinfo {author} {\bibfnamefont {I.~I.}\ \bibnamefont
  {Yakimenko}}, \ and\ \bibinfo {author} {\bibfnamefont {K.-F.}\ \bibnamefont
  {Berggren}},\ }\href {\doibase 10.1103/PhysRevB.82.073307} {\bibfield
  {journal} {\bibinfo  {journal} {Phys. Rev. B}\ }\textbf {\bibinfo {volume}
  {82}},\ \bibinfo {pages} {073307} (\bibinfo {year} {2010})}\BibitemShut
  {NoStop}%
\bibitem [{\citenamefont {Hew}\ \emph {et~al.}(2009)\citenamefont {Hew},
  \citenamefont {Thomas}, \citenamefont {Pepper}, \citenamefont {Farrer},
  \citenamefont {Anderson}, \citenamefont {Jones},\ and\ \citenamefont
  {Ritchie}}]{hew_2009_prl}%
  \BibitemOpen
  \bibfield  {author} {\bibinfo {author} {\bibfnamefont {W.~K.}\ \bibnamefont
  {Hew}}, \bibinfo {author} {\bibfnamefont {K.~J.}\ \bibnamefont {Thomas}},
  \bibinfo {author} {\bibfnamefont {M.}~\bibnamefont {Pepper}}, \bibinfo
  {author} {\bibfnamefont {I.}~\bibnamefont {Farrer}}, \bibinfo {author}
  {\bibfnamefont {D.}~\bibnamefont {Anderson}}, \bibinfo {author}
  {\bibfnamefont {G.~A.~C.}\ \bibnamefont {Jones}}, \ and\ \bibinfo {author}
  {\bibfnamefont {D.~A.}\ \bibnamefont {Ritchie}},\ }\href {\doibase
  10.1103/PhysRevLett.102.056804} {\bibfield  {journal} {\bibinfo  {journal}
  {Phys. Rev. Lett.}\ }\textbf {\bibinfo {volume} {102}},\ \bibinfo {pages}
  {056804} (\bibinfo {year} {2009})}\BibitemShut {NoStop}%
\bibitem [{\citenamefont {Smith}\ \emph {et~al.}(2009)\citenamefont {Smith},
  \citenamefont {Hew}, \citenamefont {Thomas}, \citenamefont {Pepper},
  \citenamefont {Farrer}, \citenamefont {Anderson}, \citenamefont {Jones},\
  and\ \citenamefont {Ritchie}}]{smith_2009_prb}%
  \BibitemOpen
  \bibfield  {author} {\bibinfo {author} {\bibfnamefont {L.~W.}\ \bibnamefont
  {Smith}}, \bibinfo {author} {\bibfnamefont {W.~K.}\ \bibnamefont {Hew}},
  \bibinfo {author} {\bibfnamefont {K.~J.}\ \bibnamefont {Thomas}}, \bibinfo
  {author} {\bibfnamefont {M.}~\bibnamefont {Pepper}}, \bibinfo {author}
  {\bibfnamefont {I.}~\bibnamefont {Farrer}}, \bibinfo {author} {\bibfnamefont
  {D.}~\bibnamefont {Anderson}}, \bibinfo {author} {\bibfnamefont {G.~A.~C.}\
  \bibnamefont {Jones}}, \ and\ \bibinfo {author} {\bibfnamefont {D.~A.}\
  \bibnamefont {Ritchie}},\ }\href {\doibase 10.1103/PhysRevB.80.041306}
  {\bibfield  {journal} {\bibinfo  {journal} {Phys. Rev. B}\ }\textbf {\bibinfo
  {volume} {80}},\ \bibinfo {pages} {041306} (\bibinfo {year}
  {2009})}\BibitemShut {NoStop}%
\bibitem [{\citenamefont {Birkl}\ \emph {et~al.}(1992)\citenamefont {Birkl},
  \citenamefont {Kassner},\ and\ \citenamefont {Walther}}]{birkl_1992_nature}%
  \BibitemOpen
  \bibfield  {author} {\bibinfo {author} {\bibfnamefont {G.}~\bibnamefont
  {Birkl}}, \bibinfo {author} {\bibfnamefont {S.}~\bibnamefont {Kassner}}, \
  and\ \bibinfo {author} {\bibfnamefont {H.}~\bibnamefont {Walther}},\ }\href
  {\doibase 10.1038/357310a0} {\bibfield  {journal} {\bibinfo  {journal}
  {Nature}\ }\textbf {\bibinfo {volume} {357}},\ \bibinfo {pages} {310}
  (\bibinfo {year} {1992})}\BibitemShut {NoStop}%
\bibitem [{\citenamefont {Enzer}\ \emph {et~al.}(2000)\citenamefont {Enzer},
  \citenamefont {Schauer}, \citenamefont {Gomez}, \citenamefont {Gulley},
  \citenamefont {Holzscheiter}, \citenamefont {Kwiat}, \citenamefont
  {Lamoreaux}, \citenamefont {Peterson}, \citenamefont {Sandberg},
  \citenamefont {Tupa}, \citenamefont {White}, \citenamefont {Hughes},\ and\
  \citenamefont {James}}]{enzer_2000_prl}%
  \BibitemOpen
  \bibfield  {author} {\bibinfo {author} {\bibfnamefont {D.~G.}\ \bibnamefont
  {Enzer}}, \bibinfo {author} {\bibfnamefont {M.~M.}\ \bibnamefont {Schauer}},
  \bibinfo {author} {\bibfnamefont {J.~J.}\ \bibnamefont {Gomez}}, \bibinfo
  {author} {\bibfnamefont {M.~S.}\ \bibnamefont {Gulley}}, \bibinfo {author}
  {\bibfnamefont {M.~H.}\ \bibnamefont {Holzscheiter}}, \bibinfo {author}
  {\bibfnamefont {P.~G.}\ \bibnamefont {Kwiat}}, \bibinfo {author}
  {\bibfnamefont {S.~K.}\ \bibnamefont {Lamoreaux}}, \bibinfo {author}
  {\bibfnamefont {C.~G.}\ \bibnamefont {Peterson}}, \bibinfo {author}
  {\bibfnamefont {V.~D.}\ \bibnamefont {Sandberg}}, \bibinfo {author}
  {\bibfnamefont {D.}~\bibnamefont {Tupa}}, \bibinfo {author} {\bibfnamefont
  {A.~G.}\ \bibnamefont {White}}, \bibinfo {author} {\bibfnamefont {R.~J.}\
  \bibnamefont {Hughes}}, \ and\ \bibinfo {author} {\bibfnamefont {D.~F.~V.}\
  \bibnamefont {James}},\ }\href {\doibase 10.1103/PhysRevLett.85.2466}
  {\bibfield  {journal} {\bibinfo  {journal} {Phys. Rev. Lett.}\ }\textbf
  {\bibinfo {volume} {85}},\ \bibinfo {pages} {2466} (\bibinfo {year}
  {2000})}\BibitemShut {NoStop}%
\bibitem [{\citenamefont {Shimshoni}\ \emph {et~al.}(2011)\citenamefont
  {Shimshoni}, \citenamefont {Morigi},\ and\ \citenamefont
  {Fishman}}]{shimshoni_2011_prl}%
  \BibitemOpen
  \bibfield  {author} {\bibinfo {author} {\bibfnamefont {E.}~\bibnamefont
  {Shimshoni}}, \bibinfo {author} {\bibfnamefont {G.}~\bibnamefont {Morigi}}, \
  and\ \bibinfo {author} {\bibfnamefont {S.}~\bibnamefont {Fishman}},\ }\href
  {\doibase 10.1103/PhysRevLett.106.010401} {\bibfield  {journal} {\bibinfo
  {journal} {Phys. Rev. Lett.}\ }\textbf {\bibinfo {volume} {106}},\ \bibinfo
  {pages} {010401} (\bibinfo {year} {2011})}\BibitemShut {NoStop}%
\bibitem [{\citenamefont {Gong}\ \emph {et~al.}(2010)\citenamefont {Gong},
  \citenamefont {Lin},\ and\ \citenamefont {Duan}}]{gong_2010_prl}%
  \BibitemOpen
  \bibfield  {author} {\bibinfo {author} {\bibfnamefont {Z.-X.}\ \bibnamefont
  {Gong}}, \bibinfo {author} {\bibfnamefont {G.-D.}\ \bibnamefont {Lin}}, \
  and\ \bibinfo {author} {\bibfnamefont {L.-M.}\ \bibnamefont {Duan}},\ }\href
  {\doibase 10.1103/PhysRevLett.105.265703} {\bibfield  {journal} {\bibinfo
  {journal} {Phys. Rev. Lett.}\ }\textbf {\bibinfo {volume} {105}},\ \bibinfo
  {pages} {265703} (\bibinfo {year} {2010})}\BibitemShut {NoStop}%
\bibitem [{\citenamefont {del Campo}\ \emph {et~al.}(2010)\citenamefont {del
  Campo}, \citenamefont {De~Chiara}, \citenamefont {Morigi}, \citenamefont
  {Plenio},\ and\ \citenamefont {Retzker}}]{delcampo_2010_prl}%
  \BibitemOpen
  \bibfield  {author} {\bibinfo {author} {\bibfnamefont {A.}~\bibnamefont {del
  Campo}}, \bibinfo {author} {\bibfnamefont {G.}~\bibnamefont {De~Chiara}},
  \bibinfo {author} {\bibfnamefont {G.}~\bibnamefont {Morigi}}, \bibinfo
  {author} {\bibfnamefont {M.~B.}\ \bibnamefont {Plenio}}, \ and\ \bibinfo
  {author} {\bibfnamefont {A.}~\bibnamefont {Retzker}},\ }\href {\doibase
  10.1103/PhysRevLett.105.075701} {\bibfield  {journal} {\bibinfo  {journal}
  {Phys. Rev. Lett.}\ }\textbf {\bibinfo {volume} {105}},\ \bibinfo {pages}
  {075701} (\bibinfo {year} {2010})}\BibitemShut {NoStop}%
\bibitem [{\citenamefont {Astrakharchik}\ \emph {et~al.}(2008)\citenamefont
  {Astrakharchik}, \citenamefont {Morigi}, \citenamefont {De~Chiara},\ and\
  \citenamefont {Boronat}}]{astrakharchik_2008_pra}%
  \BibitemOpen
  \bibfield  {author} {\bibinfo {author} {\bibfnamefont {G.~E.}\ \bibnamefont
  {Astrakharchik}}, \bibinfo {author} {\bibfnamefont {G.}~\bibnamefont
  {Morigi}}, \bibinfo {author} {\bibfnamefont {G.}~\bibnamefont {De~Chiara}}, \
  and\ \bibinfo {author} {\bibfnamefont {J.}~\bibnamefont {Boronat}},\ }\href
  {\doibase 10.1103/PhysRevA.78.063622} {\bibfield  {journal} {\bibinfo
  {journal} {Phys. Rev. A}\ }\textbf {\bibinfo {volume} {78}},\ \bibinfo
  {pages} {063622} (\bibinfo {year} {2008})}\BibitemShut {NoStop}%
\bibitem [{\citenamefont {Ruhman}\ \emph {et~al.}(2012)\citenamefont {Ruhman},
  \citenamefont {Dalla~Torre}, \citenamefont {Huber},\ and\ \citenamefont
  {Altman}}]{ruhman_2012_prb}%
  \BibitemOpen
  \bibfield  {author} {\bibinfo {author} {\bibfnamefont {J.}~\bibnamefont
  {Ruhman}}, \bibinfo {author} {\bibfnamefont {E.~G.}\ \bibnamefont
  {Dalla~Torre}}, \bibinfo {author} {\bibfnamefont {S.~D.}\ \bibnamefont
  {Huber}}, \ and\ \bibinfo {author} {\bibfnamefont {E.}~\bibnamefont
  {Altman}},\ }\href {\doibase 10.1103/PhysRevB.85.125121} {\bibfield
  {journal} {\bibinfo  {journal} {Phys. Rev. B}\ }\textbf {\bibinfo {volume}
  {85}},\ \bibinfo {pages} {125121} (\bibinfo {year} {2012})}\BibitemShut
  {NoStop}%
\bibitem [{\citenamefont {R\"{o}ssler}\ \emph {et~al.}(2011)\citenamefont
  {R\"{o}ssler}, \citenamefont {Baer}, \citenamefont {de~Wiljes}, \citenamefont
  {Ardelt}, \citenamefont {Ihn}, \citenamefont {Ensslin}, \citenamefont
  {Reichl},\ and\ \citenamefont {Wegscheider}}]{rossler_2011_newjphys}%
  \BibitemOpen
  \bibfield  {author} {\bibinfo {author} {\bibfnamefont {C.}~\bibnamefont
  {R\"{o}ssler}}, \bibinfo {author} {\bibfnamefont {S.}~\bibnamefont {Baer}},
  \bibinfo {author} {\bibfnamefont {E.}~\bibnamefont {de~Wiljes}}, \bibinfo
  {author} {\bibfnamefont {P.-L.}\ \bibnamefont {Ardelt}}, \bibinfo {author}
  {\bibfnamefont {T.}~\bibnamefont {Ihn}}, \bibinfo {author} {\bibfnamefont
  {K.}~\bibnamefont {Ensslin}}, \bibinfo {author} {\bibfnamefont
  {C.}~\bibnamefont {Reichl}}, \ and\ \bibinfo {author} {\bibfnamefont
  {W.}~\bibnamefont {Wegscheider}},\ }\href {\doibase
  10.1088/1367-2630/13/11/113006} {\bibfield  {journal} {\bibinfo  {journal}
  {New J. Phys.}\ }\textbf {\bibinfo {volume} {13}},\ \bibinfo {pages} {113006}
  (\bibinfo {year} {2011})}\BibitemShut {NoStop}%
\bibitem [{\citenamefont {Foulkes}\ \emph {et~al.}(2001)\citenamefont
  {Foulkes}, \citenamefont {Mitas}, \citenamefont {Needs},\ and\ \citenamefont
  {Rajagopal}}]{foulkes_2001_rmp}%
  \BibitemOpen
  \bibfield  {author} {\bibinfo {author} {\bibfnamefont {W.~M.~C.}\
  \bibnamefont {Foulkes}}, \bibinfo {author} {\bibfnamefont {L.}~\bibnamefont
  {Mitas}}, \bibinfo {author} {\bibfnamefont {R.~J.}\ \bibnamefont {Needs}}, \
  and\ \bibinfo {author} {\bibfnamefont {G.}~\bibnamefont {Rajagopal}},\ }\href
  {\doibase 10.1103/RevModPhys.73.33} {\bibfield  {journal} {\bibinfo
  {journal} {Rev. Mod. Phys.}\ }\textbf {\bibinfo {volume} {73}},\ \bibinfo
  {pages} {33–83} (\bibinfo {year} {2001})}\BibitemShut {NoStop}%
\bibitem [{\citenamefont {Umrigar}\ \emph {et~al.}(1993)\citenamefont
  {Umrigar}, \citenamefont {Nightingale},\ and\ \citenamefont
  {Runge}}]{umrigar_1993_jchemphys}%
  \BibitemOpen
  \bibfield  {author} {\bibinfo {author} {\bibfnamefont {C.~J.}\ \bibnamefont
  {Umrigar}}, \bibinfo {author} {\bibfnamefont {M.~P.}\ \bibnamefont
  {Nightingale}}, \ and\ \bibinfo {author} {\bibfnamefont {K.~J.}\ \bibnamefont
  {Runge}},\ }\href {\doibase 10.1063/1.465195} {\bibfield  {journal} {\bibinfo
   {journal} {J. Chem. Phys.}\ }\textbf {\bibinfo {volume} {99}},\ \bibinfo
  {pages} {2865} (\bibinfo {year} {1993})}\BibitemShut {NoStop}%
\bibitem [{\citenamefont {G\"u\ifmmode~\mbox{\c{c}}\else \c{c}\fi{}l\"u}\ \emph
  {et~al.}(2005)\citenamefont {G\"u\ifmmode~\mbox{\c{c}}\else \c{c}\fi{}l\"u},
  \citenamefont {Jeon}, \citenamefont {Umrigar},\ and\ \citenamefont
  {Jain}}]{guclu_2005_prb}%
  \BibitemOpen
  \bibfield  {author} {\bibinfo {author} {\bibfnamefont {A.~D.}\ \bibnamefont
  {G\"u\ifmmode~\mbox{\c{c}}\else \c{c}\fi{}l\"u}}, \bibinfo {author}
  {\bibfnamefont {G.~S.}\ \bibnamefont {Jeon}}, \bibinfo {author}
  {\bibfnamefont {C.~J.}\ \bibnamefont {Umrigar}}, \ and\ \bibinfo {author}
  {\bibfnamefont {J.~K.}\ \bibnamefont {Jain}},\ }\href {\doibase
  10.1103/PhysRevB.72.205327} {\bibfield  {journal} {\bibinfo  {journal} {Phys.
  Rev. B}\ }\textbf {\bibinfo {volume} {72}},\ \bibinfo {pages} {205327}
  (\bibinfo {year} {2005})}\BibitemShut {NoStop}%
\bibitem [{\citenamefont {Umrigar}\ and\ \citenamefont
  {Filippi}(2005)}]{umrigar_2005_prl}%
  \BibitemOpen
  \bibfield  {author} {\bibinfo {author} {\bibfnamefont {C.~J.}\ \bibnamefont
  {Umrigar}}\ and\ \bibinfo {author} {\bibfnamefont {C.}~\bibnamefont
  {Filippi}},\ }\href {\doibase 10.1103/PhysRevLett.94.150201} {\bibfield
  {journal} {\bibinfo  {journal} {Phys. Rev. Lett.}\ }\textbf {\bibinfo
  {volume} {94}},\ \bibinfo {pages} {150201} (\bibinfo {year}
  {2005})}\BibitemShut {NoStop}%
\bibitem [{\citenamefont {Umrigar}\ \emph {et~al.}(2007)\citenamefont
  {Umrigar}, \citenamefont {Toulouse}, \citenamefont {Filippi}, \citenamefont
  {Sorella},\ and\ \citenamefont {Hennig}}]{umrigar_2007_prl}%
  \BibitemOpen
  \bibfield  {author} {\bibinfo {author} {\bibfnamefont {C.~J.}\ \bibnamefont
  {Umrigar}}, \bibinfo {author} {\bibfnamefont {J.}~\bibnamefont {Toulouse}},
  \bibinfo {author} {\bibfnamefont {C.}~\bibnamefont {Filippi}}, \bibinfo
  {author} {\bibfnamefont {S.}~\bibnamefont {Sorella}}, \ and\ \bibinfo
  {author} {\bibfnamefont {R.~G.}\ \bibnamefont {Hennig}},\ }\href {\doibase
  10.1103/PhysRevLett.98.110201} {\bibfield  {journal} {\bibinfo  {journal}
  {Phys. Rev. Lett.}\ }\textbf {\bibinfo {volume} {98}},\ \bibinfo {pages}
  {110201} (\bibinfo {year} {2007})}\BibitemShut {NoStop}%
\bibitem [{\citenamefont {G\"u\c{c}l\"u}\ \emph {et~al.}(2008)\citenamefont
  {G\"u\c{c}l\"u}, \citenamefont {Ghosal}, \citenamefont {Umrigar},\ and\
  \citenamefont {Baranger}}]{guclu_2008_prb}%
  \BibitemOpen
  \bibfield  {author} {\bibinfo {author} {\bibfnamefont {A.~D.}\ \bibnamefont
  {G\"u\c{c}l\"u}}, \bibinfo {author} {\bibfnamefont {A.}~\bibnamefont
  {Ghosal}}, \bibinfo {author} {\bibfnamefont {C.~J.}\ \bibnamefont {Umrigar}},
  \ and\ \bibinfo {author} {\bibfnamefont {H.~U.}\ \bibnamefont {Baranger}},\
  }\href {\doibase 10.1103/PhysRevB.77.041301} {\bibfield  {journal} {\bibinfo
  {journal} {Phys. Rev. B}\ }\textbf {\bibinfo {volume} {77}},\ \bibinfo
  {pages} {041301} (\bibinfo {year} {2008})}\BibitemShut {NoStop}%
\bibitem [{\citenamefont {Casula}\ \emph {et~al.}(2006)\citenamefont {Casula},
  \citenamefont {Sorella},\ and\ \citenamefont {Senatore}}]{casula_2006_prb}%
  \BibitemOpen
  \bibfield  {author} {\bibinfo {author} {\bibfnamefont {M.}~\bibnamefont
  {Casula}}, \bibinfo {author} {\bibfnamefont {S.}~\bibnamefont {Sorella}}, \
  and\ \bibinfo {author} {\bibfnamefont {G.}~\bibnamefont {Senatore}},\ }\href
  {\doibase 10.1103/PhysRevB.74.245427} {\bibfield  {journal} {\bibinfo
  {journal} {Phys. Rev. B}\ }\textbf {\bibinfo {volume} {74}},\ \bibinfo
  {pages} {245427} (\bibinfo {year} {2006})}\BibitemShut {NoStop}%
\bibitem [{Note1()}]{Note1}%
  \BibitemOpen
  \bibinfo {note} {This is due to the decoupling of the two rows; at high
  densities, there are likely to be configurations where two ``adjacent''
  (ordered by their $\theta $ coordinate) electrons are in the same row, and
  the next electron is in the opposite row. (I.e., the $|i-j|=0$ and $|i-j|=1$
  electrons are in the same row, and the $|i-j|=2$ electron is in the opposite
  row, or both the $|i-j|=1$ and $|i-j|=2$ electrons are in the row opposite
  the $|i-j|=0$ electron.) Such configurations will give a negative
  contribution to $C_{zz}(|i-j|=2)$.}\BibitemShut {Stop}%
\end{thebibliography}%

\end{document}